\documentstyle[amssymb, 12pt]{article}
\newcommand{\ie}{{\rm i.e.}}
\newcommand{\beq}{\begin{equation}}
\newcommand{\eeq}{\end{equation}}
\newcommand{\bea}{\begin{eqnarray}}
\newcommand{\eea}{\end{eqnarray}}
\newcommand{\tr}{{\,\hbox{\rm Tr}\,}}
\newcommand{\la}{\left\langle}
\newcommand{\ra}{\right\rangle}

\renewcommand{\d}{{\, d}}
\newcommand{\rf}[1]{{(\ref{#1})}}
\newcommand{\hol}[3]{{{\rm Hol}_{#1}^{#2}(#3)}}
\newcommand{\sutwo}{{{\rm SU}(2)}}
\newcommand{\barx}{{\bar x}}
\newcommand{\bara}{{A_0}}
\newcommand{\lora}{\longrightarrow}

\begin{document}
\begin{titlepage}
\begin{flushright}
IFUM 493/FT  \\
hep-th/9502110
\end{flushright}
\begin{flushright}
First Version: February 1995\\
Revised Version: May 1995
\end{flushright}
\vspace{0.5cm}
\begin{center}
{\Large {\bf
Donaldson--Witten Invariants and Pure 4D-QCD\\ with
Order
and Disorder 't~Hooft-like Operators
 }} \\
\vspace{1.5cm}
{\bf Alberto S. Cattaneo}
\footnote{E-mail: cattaneo@vaxmi.mi.infn.it}  \\
\vspace{0.4cm}
{\em Dipartimento di Fisica, Universit\`a degli Studi di Milano,
20133 Milano, Italy \\
and INFN, Sezione di Milano, 20133 Milano, Italy.} \\
\vspace{0.4cm}
{\bf Paolo Cotta-Ramusino}
\footnote{
E-mail: cotta@vaxmi.mi.infn.it} \\
{\em Dipartimento di Matematica, Universit\`a degli Studi di Milano,
20133 Milano, Italy \\
and INFN, Sezione di Milano.}
\\
\vspace{0.4cm}
{\bf Andrea Gamba}
\footnote{E-mail: gamba@pol88a.polito.it} \\
\vspace{0.4cm}
{\em Dipartimento di Matematica, Politecnico di Torino, 10129 Torino,
Italy \\
and INFN, Sezione Milano.} \\
\vspace{0.4cm}
{\bf Maurizio Martellini}
\footnote{
Partially supported by EEC Science Project n. CHRX-CT93-0362.
E-mail: martellini@vaxmi.mi.infn.it} \\
\vspace{0.4cm}
{\em Dipartimento di Fisica, Universit\`a degli Studi di Milano,
20133 Milano, Italy\\
and Sezione INFN di Pavia.}
\end{center}
\vspace{2cm}

\begin{abstract}
\noindent
We study the first-order formalism of pure
four-dimensional ${\rm SU}(2)$ Yang--Mills theory with theta-term.
We describe the Green functions associated to
electric and magnetic flux operators \`a la 't~Hooft
by means of gauge-invariant
non-local operators.  These Green functions are related to Witten's
invariants of four-manifolds.
\end{abstract}
\end{titlepage}
\newpage
\baselineskip=18pt

\section{Introduction}
Seiberg and Witten \cite{SW} studied recently the strongly coupled infrared
limit of minimal
$N=2$ super Yang--Mills theory. The key point in their work is the
existence of an electric-magnetic $S$-duality \cite{VW,Gira,Ferr},
which extends the Olive--Montonen duality \cite{OM}.
By taking into account the low energy effective theory, Witten \cite{w1}
introduced an interesting
``monopole equation"\rlap,\footnote{
A monopole equation can be obtained also in the framework of the twisted $N=2$
sigma-model \cite{AF}.}
describing abelian gauge fields weakly coupled to monopoles.
By counting the
solutions of the ``dual'' equation, Witten obtains
new invariants for a compact, simply connected, oriented
4-manifold with a distinguished integral cohomology class.
These invariants are closely related to
Donaldson polynomial invariants.

On the other hand, the infrared phase of the $N=2$ super Yang--Mills
theory is like 't Hooft's ``superinsulator'' (dual superconductor),
where one gets absolute confinement of electric charges
by the monopole-condensation mechanism. This phase is
controlled by a set of topological operators, implicitly
defined by 't Hooft, that create or destroy  topological  quantum
numbers.
We call such operators  ``disorder  operators''.
't~Hooft  argued
that the correlators of the disorder operators  are  a  sort  of
``topological'' Green's function.

In~\cite{cm,ccm,c}
the authors  consider three- and four-dimensional
$BF$-theory and define new
gauge-invariant, non-local operators $M(\Sigma)$ associated to
surfaces $\Sigma$ immersed or embedded in a  four-dimensional  manifold.

It turns  out  that
these operators $M(\Sigma)$ are  an  explicit  realization  of
't~Hooft disorder operators.

In this work we generalize the ideas of
\cite{cm,c,ccfm} and
prove that the first-order formalism of the  standard  pure
bosonic Yang--Mills theory with theta-term (which is a special type
of $BF$-theory) allows us to get
an explicit  realization  of  the  't~Hooft
scenario.

Moreover we show that the  expectation
values  of  the  non-local  operators $M(\Sigma)$ in the weak coupling
regime provide solutions of a ``monopole equation".

When we consider a K\"ahler manifold, then our monopole
equation is related to the one considered by Witten,
provided that some extra relations are fulfilled \cite{w1} (see
\rf{spinor}).

The inclusion in the expectation values of
$M(\Sigma)$ of the ``order parameters'' provided
by Wilson operators ``perturbs'' the above topological
Green functions. The new  correlation functions appear to
be related to Donaldson invariants.

We would like to stress that the main ingredient
here
is the fact that monopole equations select singular abelian
connections. In our picture the
``source" of the monopole equation is provided by a bosonic 2-form,
which appears naturally in the first order formalism of
pure Yang--Mills theory, while in the $N=2$ super Yang--Mills theory \cite{w1}
such a source is given by a bilinear form of a complex Weyl spinor.

Our approach may give a hint on how to perform a
completely non-perturbative calculation  directly  in
standard 4D-QCD, without the need to get it by some spontaneous
symmetry breaking of an $N=2$ super Yang--Mills theory \cite{SW}.
Our result gives more support to the general intuition that in
standard 4D Yang--Mills theory, there should be a  topological  sector
controlled by the topological Green's functions of  suitable
gauge-invariant, non-local, composite operators.

\section{4D-QCD in first order formalism}

Let us start with some mathematical definitions.
Let $X$ be a compact, oriented 4-dimensional  manifold,  and
more specifically a K\"ahler  manifold  with  K\"ahler  form
$\omega$ (i.e. $\omega$ is a closed $(1,1)$-form).
On $X$ we consider a non-abelian gauge theory.
{}From a mathematical point of  view  this  amounts  to  defining  a
complex vector bundle $E$ associated to a principal bundle
with structure group $G$.
We will always consider the case of $G={\rm SU}(2)$.
We will write ${\frak g}_E$  for  the  bundle  associated  to  the
adjoint representation of ${\rm SU}(2)$.
If $A$ denotes a unitary connection on $E$, then
we denote with
$d_A:\Omega_X^p(E)\rightarrow\Omega_X^{p+1}(E)$
the
associated covariant exterior derivative.
Notice that $d_A=d+A$ on $\Omega_X^0(E)$.
Here $A$ stands for an ${\frak su}(2)$-valued one-form
which transforms locally as
$A\rightarrow A_u\equiv u^{-1}Au+u^{-1}du$,
where $u$ is (a local representative of)
an element of the gauge group, \ie\ a smooth map
$u:X\rightarrow {\rm SU}(2)$,
and $\Omega_X^p(E)$
denotes as usual
the space of $p$-forms on $X$ with values in $E$.
The Yang--Mills curvature, i.e. the  curvature  of  the  connection
$A$, is the ${\frak su}(2)$-valued two-form
$F_A=dA+A\wedge A\in \Omega_X^2({\frak g}_E)$
which transforms as
$F_A\rightarrow F_{A_u}\equiv u^{-1}F_A u$.
Notice that for any connection $A$ we have the Bianchi identities
$d_A F_A=0$.
If we fix a  metric  structure  on  $X$  and  denote  by  $*$  the
associated Hodge operator we have the splitting into self-dual and
anti-self-dual parts of the bundle-valued 2-form
\beq\label{tre}
F_A=F_A^+\oplus F_A^-\in\Omega_X^{2,+}({\frak g}_ E)\oplus
                         \Omega_X^{2,-}({\frak g}_ E)
\eeq
We call the connection $A$   anti-self-dual (ASD) if
$F_A^+\equiv \displaystyle{\left({1+*\over 2}\right)}F_A=0$.
Since $X$ is by  hypothesis  a  Hermitian  manifold  we  can  also
decompose the curvature $F_A$ as
\beq\label{quattro}
F_A=F_A^{(2,0)}\oplus F_A^{(1,1)}\oplus F_A^{(0,2)}
\in\Omega_X^{(2,0)}({\frak g}_E)\oplus\Omega_X^{(1,1)}({\frak g}_E)
\otimes\Omega_X^{(0,2)}({\frak g}_E).
\eeq
Now one can show that  the  self-dual  bundle-valued  complexified
2-forms over $X$ are
\cite{dk}:
\beq\label{cinque}
\Omega_X^{2,+}({\frak g}_E)
=\Omega_X^{(2,0)}({\frak g}_E)\oplus\omega\Omega_X^0({\frak g}_E)
\oplus\Omega_X^{(0,2)}({\frak g}_E)
\eeq
with $\omega$ the K\"ahler $(1,1)$-form
($\omega\Omega_X^0({\frak g}_E)=\Omega_X^{(1,1)}({\frak g}_E)$).
In~(\ref{tre}-\ref{cinque}) the
$\Omega_X^{(p.q)}({\frak g}_E)$
denote the sections
$\Gamma(\Lambda_X^{(p,q)}\otimes{\frak g}_E)$
and the $\Omega_X^\pm$
stand for the algebraic projections
$\displaystyle{{1\over 2}(1\pm *)}\Omega_X\equiv \pi_\pm\Omega_X$.

Our physical theory
is the first-order version of the pure  ${\rm SU}(2)$ Yang--Mills
theory with the so-called ``theta-term'' given by the action
functional (in Euclidean signature)
\beq\label{sei}
S_{BF^+}=\int_X{\rm Tr}(B\wedge F_A^+)
-{g^2\over 4}\int_X{\rm Tr}(B\wedge B),
\eeq
$B\in\Omega_X^2({\frak g}_E)$,
and hence $B\rightarrow B_u\equiv u^{-1}B u$
under gauge transformations.
In~\rf{sei}, $g^2$ denotes the bare gauge coupling constant.
The field equations obtained from~\rf{sei} are
\bea
\label{settea}
F_A^+&=&{g^2\over 2}B, \\
\label{setteb}
(d_A^+)^*B&=&0
\eea
where  the $+$ superscript  denotes  the  self-dual
projection and
$d_A^*:\Omega_X^{p+1}(E)\rightarrow \Omega_X^p(E)$
is the
adjoint operator of the covariant derivative.
Eq.~\rf{settea} clearly implies that $B\in\Omega^{2,+}$
and hence one may set $B=B^+$.
In the following we shall denote by ${\cal M}_c$
the ``classical'' moduli space, that is the space of
solutions  of~(\ref{settea}-\ref{setteb})  modulo  the
gauge group.
Now  using~\rf{settea}  in~\rf{sei}  one   gets   the   action
functional  in  the  second-order  formalism  (in  the   Euclidean
signature):
\beq\label{nove}
S_{\rm YM}={1\over g^2}\int_X{\rm Tr}(F_A^+\wedge F_A^+)
={1\over 2g^2}\int_X{\rm Tr}(F_A\wedge *F_A)+
{1\over 2g^2}\int_X{\rm Tr}(F_A\wedge F_A).
\eeq
At this point one may reabsorb the gauge coupling $g$  into  $F_A$
through the rescaling $A\rightarrow A'=A/g$, so that \rf{settea} now reads
\beq\label{Fprimo}
F_{A'}^{\prime +} \equiv (dA'+gA'\wedge A')^+ = \frac g2 B.
\eeq

Notice that the first term in the right-hand-side of~\rf{nove}
is the standard Yang--Mills functional, while the second one
is the topological charge, the so-called ``theta-term''
associated with the instanton number $k$
\cite{dk}
\beq\label{dieci}
k(A)=-c_2(E)[X]=-{1\over 8\pi^2}\int_X{\rm Tr}(F_A\wedge F_A).
\eeq
Here $c_2(E)$ is the second Chern class of the complex
${\rm SU}(2)$-vector bundle $E$ and we have adopted the standard sign
convention for $k$.
If we denote by $t_a$ the generators of ${\frak su}(2)$ (in the fundamental
representation) we can write
$A=\sum_a A_\mu^a t_a dx^\mu$,
$F_A=\sum_a(F_A)_{\mu\nu}^at_adx^\mu\bigwedge dx^\nu$
in a local coordinate system $\{x^\mu\}$ on $X$.
Furthermore, we fix the normalization for the $t_a$'s according
to
${\rm Tr}(t_at_b)=-{1\over 2}\delta^{ab}$.
For ${\rm SU}(2)$ we get that $t_a=i\sigma_a/2$ $(a=1,2,3)$
with $\sigma^a$ the Pauli matrices.

In the following we shall consider the quantum  theory  associated
to  the  action~\rf{sei}  in  the  frame  of  path-integral
quantization.
For  this  purpose   we   need   to   fix   the   gauge-invariance
of~\rf{sei}.
We use the background covariant ``gauge fixing" condition
\beq\label{dodici}
d_{A_0}^*a=0=d_{A_0}^*B,
\eeq
where we have defined $A=A_0+ a$ with $A_0$ a fixed background
${\rm SU}(2)$-connection.
Notice that we have to impose a constraint on the field $B$, since the theory
acquires on shell a larger symmetry. We will stick to the common
terminology and call this constraint on the field $B$ a ``gauge fixing".
Since~\rf{sei} is just the 4-dimensional Yang--Mills theory
(with theta-term)
\footnote{
In Euclidean signature the
Gibbs measure associated to the
``theta  action-functional''is  usually
written as $\exp(i\theta k(A))$, where $k(A)$ is defined by~\rf{dieci}
and the Lagrange multiplier $\theta$ is the so-called $\theta$-parameter.
Then~\rf{nove} implies
$\theta=4\pi^2$, which is consistent with the fact that
$\theta$ does not  renormalize.
}
written in the first-order formalism, and we know that
a non-abelian gauge theory is asymptotically free, it follows that
the $B$ field is weakly (strongly) coupled to the Yang--Mills
potential $A$ in the ultraviolet (infrared) regime.
Therefore~\rf{Fprimo}, after the replacing of $g$
with the running coupling constant $g(\mu)$
($g(\mu)\sim 0$ at energies $\mu\gg M$ with respect to
some physical energy scale $M$)
becomes the ASD-condition $F_{A'}^{\prime +}\sim 0$
in the ultraviolet regime.

An alternative theory which, in the semiclassical approximation,
is equivalent to the one described by \rf{sei}
is given by the following topological action
\beq
S_{BF}=\int_X{\rm Tr}(B\wedge F_A)
-{g^2\over 4}\int_X{\rm Tr}(B\wedge B),
\label{BF}
\eeq
provided that we assume:
\beq\label{singular}
B^-=0.
\eeq
By assuming the above constraint, we are breaking the ``topological"
symmetries of the $BF$ theory \rf{BF} given by:
\bea\label{topo}
\delta_t A &=& \frac{g^2}2 \eta\\
\delta_t B &=& -d_A \eta,
\eea
where $\eta$ is a 1-form ghost \cite{cm}\cite{c}.

\section{Quantum theory}

The quantum theory associated to~\rf{sei} is determined  by  the
functional integral
\beq\label{tredici}
\la 1 \ra_{\cal M}
=\int_{\cal M} d\mu(A,B)
e^{-S_{BF^+}}.
\eeq
In~\rf{tredici} $\cal M$ is the orbit space, i.e. the  quotient
of the set of fields $(A,B)$ by the gauge group.
Following the usual procedure, the  functional measure  $d\mu(A,B)$  over
$\cal M$ is  obtained  by  the  standard  Faddeev-Popov  procedure
\beq\label{quattordici}
\left. d\mu(A,B)\right|_{A=A_0+a}
=[da][dB]\delta(d_{A_0}^*a)\delta(d_{A_0}^*B)J(A,B),
\eeq
with $a\in\Omega^1({\frak g}_E)$ and
the functional Jacobian
$J(A,B)=\det'[\delta_\mu(d_{A_0}^*a_u)]
\det'[\delta(d_{A_0}^*B_u)]$
is the standard Faddeev-Popov determinant.
Of course, in  order
to derive~\rf{quattordici} in the background gauge  one  assumes
that the topology of $X$ allows only  irreducible
connections  $A$  on  the  ${\rm  SU}(2)$-bundle  $E$ over $X$.
This means
that there are no covariantly constant scalar fields.
A sufficient condition for that~\cite{dk} is  that
$b_2^+>1$; here $b_2^+$ is the dimension of  the  space
of self-dual harmonic forms over $X$.

Our  key  idea is  to  include   into~\rf{tredici}   some
gauge-invariant   non-local   operator $M(\Sigma)$
(to be defined in the next section),
which is
analogous to the disorder loop variable of 't~Hooft
\cite{th1},
in such a way that one has
\beq\label{quindici}
\la M(\Sigma) \ra_{\cal M}
=\int_{\tilde{\cal M}} d\tilde\mu_{L_\Sigma}(\tilde A,\tilde B)
e^{-\tilde S_{BF^+}}
\equiv\la 1 \ra_{(\tilde {\cal M},L_\Sigma)}
\eeq
where $\tilde{\cal M}$ is a particular case (namely the 0-dimensional
one) of the  Witten
moduli  space  which  gives   the   new   differential-topological
invariants   for   compact,      4-manifolds    recently
discovered by
Witten~\cite{w1}.
Eq.~\rf{quindici} is one of our basic results.
In the following we shall   define  explicitly  in  terms  of
$(A,B)$  a  non-local   gauge   invariant   operator   $M(\Sigma)$
associated   to   a   surface   $\Sigma\subset   X$,   such
that\rf{quindici} holds identically, in the semiclassical
approximation and in the weak coupling limit (up to some  irrelevant
normalization).
In~\rf{quindici} the symbols have the following meaning: the introduction
of the observable $M(\Sigma)$ brings into the picture (singular)
reducible connections that are not allowed in the theory without such
observables. In other words we obtain (abelian) connections
(denote by the symbol $\tilde A$)
on the holomorphic line bundle $L_\Sigma$
over $X$ having first Chern class $c_1(L_\Sigma)$ equal  to  the
K\"ahler class $[\omega]\in H^2(X)$ which is the  Poincar\'e  dual
(PD) of [the homology class of] the surface  $\Sigma\subset  X$, i.e.
$[\Sigma] ={\rm
PD}([\omega])={\rm PD}(c_1(L_\Sigma))$\footnote{From now on we will
drop the square parentheses denoting  (co)homology classes.}.
$\tilde B$ is the 2-form $B$ restricted to a  generic  $
(1,1)$-section of the bundle
$\Lambda^{(1,1),+}_X\otimes L_\Sigma$ over $X$.
Notice that the condition that $\Sigma$ be Poincar\'e dual to
the K\"ahler form $\omega$ means that there is a  singular  1-form
$\eta$ on $X$ such that for any 2-form $\hat\theta$ over  $X$  one
has
\beq\label{sedici}
\Sigma={\rm PD}(\omega)\iff
\int_{\Sigma\subset X}\hat\theta-\int_X\hat\theta\wedge\omega
=\int_Xd\hat\theta\wedge\eta.
\eeq
In~\rf{quindici} the action functional
$\tilde S\equiv S_{BF^+}(\tilde A,\tilde B;L_\Sigma;X)$
is the restriction of~\rf{sei} to the fields
$(\tilde A,\tilde B)$ defined above.
Equivalently one can say that the introduction of the operator
$M(\Sigma)$
into the expectation value~\rf{tredici}
gives an {\em effective} gauge theory which is controlled by a
 split $\>{\rm SU}(2)$-bundle
$E_\Sigma=L_\Sigma\oplus L_\Sigma^{-1}$,
where $\Sigma\rightarrow L_\Sigma$ by the correspondence
$\Sigma={\rm PD}(\omega)={\rm PD}(c_1(L_\Sigma))$.

\section{'t Hooft-like disorder operators}

In a four-dimensional  $BF$-theory, we can define
as in ref.~\cite{cm}, a   non-local   gauge   invariant
operator associated to any closed oriented immersed surface $\Sigma$:
\bea
\label{diciassette}
{O}(A,B;\Sigma,\gamma,\gamma',\bar x)
&\equiv& 2\pi q_m g\int_{\Sigma}
{\rm Tr}_R\left[
{\rm Hol}_{\bar x}^{y}(\gamma) B(y) {\rm  Hol}_{y}^{\bar  x}(\gamma')
\right]    \\
\nonumber
B(y)&\equiv&\sum_a B^a(y)R_a\equiv\sum_a B_{\mu\nu}^a(y)
R^ady^\mu\wedge dy^\nu
\eea
where $q_m$ will play the r\^ole of a ``magnetic charge"
(as it will be clear after~\rf{phim}),
$\bar   x$   is   a   base   point   on
$\Sigma$
(equivalently, one may remove the point $\bar x$ from $\Sigma$),
$R^a$  is  an  irreducible  representation\footnote{We will consider
later on, only the fundamental representation.}  of
${\rm SU}(2)$,  ${\rm  Tr}_R$  denotes  the  trace  over  $R$ and
$\gamma'\cup\gamma$ is any closed path $C$ with base point
$\bar x$ and passing through $y$.

Furthermore, $\hol{\bar x}{y}{\gamma}$ in~\rf{diciassette}
stands for the holonomy of the $\sutwo$-connection $A$ along the
path $\gamma$ from $\barx$ to $y$:
\beq\label{venti}
\hol{\barx}{y}{\gamma}
\equiv P\exp(\int_\barx^y A(x))
\eeq
It has been shown in~\cite{cm,c} that the choice of the base point $\barx$
is irrelevant\footnote{
Of course the expectation values of the operators $O$ depend
on the choice of the paths $\gamma,\gamma'$, but this choice is irrelevant
for the computation in this framework of the Donaldson-Witten
invariants.},
and underlying a choice for the
closed path $C$, one can finally set
in this context:
\beq\label{ventuno}
M(\Sigma)\equiv\exp[O_+(\Sigma)], \qquad
O_+(\Sigma)\equiv O[A,B=B^+;\Sigma,C,\bar x],
\eeq
where $O$ is defined by~\rf{diciassette}. Here of course
$O_+$ is the projection by ${1\over 2}(1+*)$ of
the integrand of $O$
to the self-dual one. We also have to consider the standard Wilson loop
observable
\beq
W(C) \equiv \tr_R\hol{}{}{C}
\eeq
where $C$ is a closed loop.

Our next step is to compute
the semiclassical approximation of
$\la M(\Sigma)\ra_{\cal M}$.
We now consider a background {\em quasi-ASD connection $A_0$},
namely a connection satisfying eq.\rf{settea} and \rf{setteb} with $B_0\neq 0$
and $g$ small enough but different from $0$.
The linear version of
the field equations~(\ref{settea},\ref{setteb})
yields the
the following equations for the field fluctuations $A_0\lora
A_0+a$ and $B_0\lora B_0 +b$ with $(a,b)\in
\Omega^1\oplus \Omega^{2,+}.$
\bea
\label{ventiduea}
d_\bara^+a &=& {g^2\over 2}b   \\
\label{ventidueb}
(d_\bara^+)^*b &=& 0.
\eea

The operator $t_g$ corresponding to~\rf{ventiduea},
\rf{ventidueb} defined by
\bea\nonumber
t_g:\Omega^1\oplus\Omega^{2,+} &\mapsto & \Omega^1\oplus\Omega^{2,+}  \\
\label{deft}
(a,b) &\rightarrow& (({d_{A_0}^+})^*b,d_{A_0}^+a-{g^2\over 2}b).
\eea
Pure gauge fluctuations $\phi\in \Omega^0$
due to the gauge transformations of the backgrounds $(A_0,B_0)$
are not relevant here
since by choosing:
$a=d_{A_0}\phi$ and $b=[B_0,\phi]$ we have that the condition
$(d_{A_0}^+)^*[B_0,\phi]=0$ is compatible with the equation
\[
\left(d^+_{A_0}d_{A_0}\phi, [B_0,\phi]\right)=
{g^2\over 2}\parallel [B_0,\phi]\parallel ^2
\]
only if we set $\phi=0$.

Hence the dimension of our moduli space will be simply given by the dimension
of $\ker t_g$.
By using the Weitzenb\"ock formulae it is possible to show
that $(a,b)\in\ker t_g$ implies $b=0$.
The same argument should also imply that
$(d_{A_0}^+)^*d_{A_0}^+ a=0$ holds only when $a=0$, provided that $A_0$ is
a quasi-ASD (not ASD) connection.
 %\beq\label{complex}
%0\rightarrow
%\Omega^0 \stackrel{s}{\rightarrow}
%\Omega^0 \oplus
%\Omega^1\oplus\Omega^{2,+} \stackrel{t_g\oplus d_\bara}{\rightarrow}
%\Omega^0\oplus\Omega^1\oplus\Omega^{2,+}\rightarrow 0
%\eeq
%In the above complex $h_0=\dim \ker s=0$ since $A_0$ is an irreducible
%connection.
%A direct computation shows that,
%as long as we keep $g\ne 0$, we have $h_2=\dim\ker (t_g\oplus d_\bara)^*=0$.

The vanishing of $\ker t_g$
can be restated by saying that the classical moduli space ${\cal M}_c$ is
zero-dimensional
or discrete.
In other words the inclusion
of the field $B$
(with $g\ne 0$) seems to remove the degeneracy of the instanton
vacua of the classical Yang--Mills theory.
We will essentially find points of the classical moduli
space, given by {\em singular} solutions.
In the semiclassical computation of
$\la M(\Sigma)\ra_{\cal M}$
one may set
${\cal M}\simeq{\cal M}_c$.

{}From the point of view of the field equations, the insertion into
the path-integral~\rf{tredici}
of the operator $M(\Sigma)$ is equivalent to considering
the improved gauge-fixed $BF^+$-action
\beq\label{ventiquattro}
S_{BF^+}(A,B,d_\bara^*A=0=d_\bara^*B;X)+O_+(\Sigma)
\equiv S_{BF^+}^{(\Sigma)}(A,B,d_\bara^*A=0=d_\bara^*B;X,\Sigma\subset X)
\eeq
Here $O_+(\Sigma)$ plays the r\^ole of a ``source term'' for~\rf{sei}. Indeed,
when we make the following choice of the
2-form $\hat\theta$ in~\rf{sedici}
\beq\label{venticinque}
\hat\theta(\Sigma)\equiv 2\pi q_mg\tr_R[\hol{\barx}{y}{\gamma}
B^+(y)\hol{y}{\barx}{\gamma'}],
\qquad
O_+(\Sigma)=\int_\Sigma\hat\theta(\Sigma)
\eeq
we have:
\beq\label{ventisei}
\int_{\Sigma}\hat\theta=\int_X\omega\wedge\hat\theta,
\qquad\omega\equiv {\rm PD}(\Sigma)
\eeq
Eq.~\rf{ventisei} follows from the fact that $\theta$ is a closed form. In fact
we have:
\bea
\label{ventisette}
d\hat\theta(\Sigma)&=& 2\pi q_mg
d\tr_R[\hol{\barx}{y}{\gamma}B^+(y)\hol{y}{\barx}{\gamma'}]  \\
&=&2\pi q_mg\tr_R
[\hol{\barx}{y}{\gamma}d_\bara B^+(y)\hol{y}{\barx}{\gamma'}]=0
\nonumber
\eea
Notice that in~\rf{ventisette} we have used the gauge condition
$d_\bara^* B=0$ which is the same as $*d_{A_0}*B=0$, and implies $\d_{\bara}
B=0$
since $B$ is self-dual.

\section{Moduli space and topological invariants}
\label{sezquattro}

In this section we prove~\rf{quindici} in the semiclassical
approximation. The starting point is given by the
field  equations obtained
after the inclusion
of  a disorder
operator  $M(\Sigma)$ \`a la 't~Hooft.

These field equations are obtained
by applying the functional derivative
$\displaystyle{{ \delta\over{\delta B^a}}}$ to
the action $S^{(\Sigma)}_{BF^+}$
(see (\ref{ventiquattro}-\ref{ventisei})) and read:
\bea
\nonumber
F_+^{(2,0),a}&=&0+O(g^2)=F_+^{(0,2),a}  \\
\label{ventotto}
F_+^{(1,1),a}(x)&=&-2\pi q_mg\omega(x){\rm Tr}_R
\{\hol{\barx}{x}{\gamma}R^a\hol{x}{\barx}{\gamma'}\}
+O(g^2)=  \\
\nonumber
&=&- 2\pi q_mg\omega(x){\rm Tr}_R\{\hol{}{}{C_x}
R^a\}
+O(g^2).
\eea
Here $C_x$ is
a loop based at $x$,
$\{R^a\}$    is  the basis   of a
representation of
${\rm SU}(2)$,
 $F_+^{(p,q)}$ is the $(p,q)$-part of $F^+$
and the other notation is the same as in the previous section.
The holonomy is non trivial if we think that
the loop $C_x$ includes a ``Dirac singularity"
(see below).
{}From now on we choose the fundamental
representation of $\sutwo$ and hence we set $R_a\equiv t_a; a=1,2,3$.
As a consequence of equation \rf{ventotto}
the curvature $F_A$ is given, in the limit $g\to 0$, by
the product of a numerical constant times
$\omega$ times the projection of ${\rm Hol}(C)$ into the
Lie algebra of
$SU(2)$. This projection defines a constant direction
in the Lie Algebra, so  we can assume that this
direction is parallel to $t_3.$
This is equivalent to considering a (singular) reducible connection.
We can furthermore choose $\omega$
as a {\em self-dual 2-form}. In conclusion
Eq.~\rf{ventotto} admits a non-trivial  singular reducible\footnote
{
In order to implement the Faddeev-Popov
procedure, we required that
there are no {\em smooth} reducible connections on $X$.
There is no contradiction here, since $\tilde A$ is singular on $\Sigma$.
}
quasi ASD
connection of the form~\cite{fu}:
\beq
\label{ventinovea}
\tilde A(x)={i\over 2}
\left(
\begin{array}{cc}
\alpha(x) & 0 \\ 0 & -\alpha(x)
\end{array}
\right)
=\alpha(x)t_3,
\eeq
where $\alpha$ is a 1-form,
$t_3=\frac{i}2\sigma_3=\frac{i}2{\rm diag}(1,-1)$.
Hence $\alpha$ is an {\em abelian} connection
and the curvature of \rf{ventinovea} is
$ft_3$,   where we have set:
$f\equiv d\alpha$.
Thus
the  introduction   of   the   operators
$M(\Sigma)$ replaces the original non-abelian Yang--Mills
theory [with theta-term]
to an effective theory based on a split $SU(2)$-bundle
$E_\Sigma=L_\Sigma\oplus  L_\Sigma^{-1}$,  with  $L_\Sigma$  a
holomorphic line bundle on $X$  ``parameterized'' by  $\Sigma\subset
X$, in the sense that $c_1(L_\Sigma)=\omega={\rm PD}(\Sigma).$
So the instanton number
$k=-c_2(E_\Sigma)$  is expressed in terms of
the monopole number $\lambda=c_1(L_\Sigma)$, since one has
$k=-c_2(E_\Sigma= L_\Sigma\oplus L_\Sigma^{-1})=c_1(L_\Sigma)^2=
\lambda^2$.
As a consequence of~\rf{ventinovea}, we have:
\beq
\label{trentaa}
{\rm Tr}(\hol{}{}{C}t_a)=-\sin (g\phi_m/2) \delta_{a,3}
\eeq
where
$\phi_m$ is the monopole magnetic flux of  $\tilde F/g$
across  the  ``Dirac  surface''  $\Sigma$.
One may regard the r.h.s.\ of~\rf{ventotto} as an effective field
$\tilde B$, given by
\beq\label{Btilde}
\tilde B^a=-\pi gq_m \omega {\rm Tr}({\rm Hol}(C)t_3)\delta_{a,3}
\eeq
obtained by the insertion of the operator $M(\Sigma)$
into~\rf{tredici} in the limits $g\rightarrow 0$ and
$\hbar\rightarrow 0$. Notice that by \rf{Dirac} the field $\tilde B$ is
actually independent of $g$ and hence contributes also in the weak-coupling
limit.
By  \rf{ventotto} and \rf{Btilde},
the self dual part of the curvature of
the vector potential $\tilde A'=\tilde A/g$
is given by:
\beq
\label{nuova}
d^+ \tilde A'=(2/g)\tilde B+O(g^2)
\eeq
Eqn. \rf{nuova} is the same as \rf{Fprimo} when applied
to the reducible connection $\tilde A$, provided that we perform
the so called weak-strong coupling duality $g\to 1/g.$
Furthermore~\rf{ventidueb} becomes
%the twisted K\"ahler-Dirac equation associated to the
%abelian connection with monopoles \rf{ventinovea}, i.e.
\beq\label{trentab}
(d^+_{\tilde A})^*\tilde B=0+O(g)
\eeq
where
$\tilde B=\tilde B^+$
 is a
$(1,1)$-section of the twisted self-dual
vector bundle $\Lambda^{(1,1),+}_X\otimes L_\Sigma$ over $X$.
Equations~(\ref{nuova}-\ref{trentab})  are the
monopole equations that, in the case of  a
general K\"ahler manifold $X$, correspond to Witten's monopole
equations \cite{w1}.

The relation between the complex Weyl spinors $M$ and $\bar M$
in Witten's equation and our
field $\tilde B$ is of the form:
\beq\label{spinor}
\tilde B_{\mu\nu} = \frac i2 \bar M\Gamma_{\mu\nu} M
\eeq
where $\Gamma_{\mu\nu}=\frac12[\Gamma_\mu,\Gamma_\nu]$ and the $\Gamma$'s are
the Clifford matrices. In the general case the solutions of our monopole
equations \rf{nuova} and \rf{trentab} are only a subclass of Witten's.
The situation here is similar to the comparison between a four-dimensional
$BF$ theory to a gravitational theory in the first order formalism.
In order to connect the above two theories one has to require that the
$B$ field is the ``square" of the vierbein \cite{js}. In the situation
described
in this paper, in order to connect our $BF$ theory with Witten
monopole equation, one has to assume the validity  of \rf{spinor}.

However,
we want to stress that the results we present in the following of this
paper are completely independent on the above
relation between our and Witten's monopole
equations.

Let us now look for an explicit  solution of~\rf{nuova}.
Requiring that, in the limit $g\to 0$, we have non trivial
solutions of \rf{ventotto}
%The condition that  there  are {\it no}
%flat abelian instantons
implies $\sin({g\over 2}\phi_m)\ne 0$.
In the path integral the effect of the cosmological-like term
\beq\label{cosm}
\Delta S=-{g^2\over 4}\int_X \tr(B\wedge B)
\eeq
with $g\approx 0$ but $g\ne 0$
is to select
the minimal energy solutions, obtained
when $|\sin({g\over 2}\phi_m)|=1$, thus giving flux quantization.

We can choose coordinates $(x,y,u,v)$ on $X$ locally in a
neighborhood of the base point $\bar x^\mu$ so that the
surface $\Sigma$ is given by the equations $x=y=0$.
For the Poincar\'e dual form $\omega$ restricted to $\Sigma$ we can
take~\cite{dk}
\beq\label{trentacinquez}
\omega(x,y)=\psi(x,y)dx\wedge dy
\eeq
where $\psi$ is a bump function on ${\bf R}^2$, supported
near $\vec{\bar x}=(x,y)=(0,0)$ and with integral $1$.
We can choose
$\psi(x,y)=\delta^{(2)}(x,y)$, with $\delta^{(2)}(x,y)$
the two-dimensional delta function.
Therefore we may think that the Maxwell's field strength $f$
is not zero only on $\Sigma$ by choosing
$\alpha=\alpha_\mu dx^\mu=\alpha_1 dx+\alpha_2 dy$,
i.e. $\alpha_3=\alpha_4=0$, so that
$f=d\alpha=f_{12}dx\wedge dy$,
where
$f_{12}\equiv(\partial_1\alpha_2-\partial_2\alpha_1)$,
$\partial_i\equiv\partial/\partial x^i$ ($i=1,2$),
$\vec x\equiv\{x^i\}\equiv(x,y)$.
Then~\rf{nuova} through~\rf{trentacinquez} becomes at order $g^2$
\beq\label{trentaseiz}
f_{12}(x,y)=\mp 2\pi q_mg
\delta^{(2)}(x,y)
\eeq
The Maxwell potential $\alpha_i$ which solves~\rf{trentaseiz}
can be written as\footnote{
Dropping the assumption of minimal energy, the general solution
is
$\alpha_i(\vec x)=cg(\epsilon_{ij}+i\delta_{ij})
\partial^j\log|\vec x-\vec {\bar x}|$
with $c/q_m=\sin (\pi g c)$.
}
\beq
%\nonumber
\alpha_i(\vec x)=\pm  q_mg(\epsilon_{ij}+i\delta_{ij})
\partial^j\log|\vec x-\vec {\bar x}|
\label{trentasettez}
\eeq
This is the Kohno connection \cite{Koh} that is obtained in a
similar contest also in \cite{msz}.
Indeed one has that
\beq\nonumber
\partial_i\partial^i\log|\vec x-\vec{\bar x}|
=2\pi\delta^{(2)}(\vec x-\vec{\bar x})
\eeq
and locally around $\vec{\bar x}$ the monopole magnetic flux
of $\alpha/g$ is
\beq\label{phim}
\phi_m=\pm 2\pi q_mQ(\Sigma,\Sigma)
\eeq
where
\beq\label{intersec}
Q(\Sigma,\Sigma')=\int_X\omega[\Sigma]\wedge\omega[\Sigma']
\eeq
denotes the {\it algebraic intersection number}~\cite{dk}
of the oriented surfaces $\Sigma$ and $\Sigma'$.
Notice that~\rf{intersec} is well defined even when
$\Sigma'=\Sigma$.
For the minimal energy solution one has $\sin (g\pi q_m)=1$,
implying
\beq\label{Dirac}
2g q_m Q(\Sigma,\Sigma)\equiv 1\quad({\rm mod}\, 4)
\eeq
which corresponds to the {\it Dirac quantization condition}.
{}From~\rf{phim} it follows that the operator~\rf{ventuno}, when expressed
in terms of the fields $(\tilde A$, $\tilde B)$ is related to
the exponential of the ``magnetic flux".

Summarizing,  we   have   shown   that   the   inclusion  of   the
gauge-invariant non-local operator
$M(\Sigma)$ into the expectation value~\rf{tredici}  given
by   the   non-abelian   ${\rm    SU}(2)$    $BF^+$-theory
reduces  this theory   to   an  { effective   abelian} one {
with monopoles} and
{\it without} the $M(\Sigma)$ operator.
The resulting quantum field theory, which is ``dual" to the original one,
is coupled to monopoles and is
written in terms of an abelian connection $\alpha$
on  a  holomorphic
line bundle $L_\Sigma$ (the monopole line bundle),
and  a
self-dual
two-form
$\tilde B$.
This result proves~\rf{quindici}
in the  semiclassical  approximation.

At  this  perturbative  order
the underlying moduli space is the classical one
and coincides with the moduli
space $\tilde {\cal M}$.

Thus the partition function given by the path-integral written  in
the r.h.s. of~\rf{quindici}  (or  equivalently  the  ``one-point''
function described by the l.h.s. of~\rf{quindici}) is equal to the
number of points in our moduli-space of the fields
$(\tilde A, \tilde B)$,
counted with a $\pm$ sign
\beq\label{trentadue}
\la 1\ra_{(\tilde{\cal M},L_\Sigma)}\equiv\#{\tilde{\cal M}}
(\tilde A,\tilde B; X,L_\Sigma)\quad   .
\eeq
Equation~\rf{trentadue} is  the same as  in
Witten~\cite{w1,w3}.
Equation ~\rf{trentadue} follows from the fact  that
the integration over $\tilde B$, gives
a delta-functional contribution,
i.e. $\delta(d^+\tilde A +2\pi q_mg\omega{\rm Tr}\{\hol{}{}{C}
t^3\})$,
which in turn gives a counting measure (with signs) on
the space $\cal M$.

Observe that the $\pm$ signs are provided by the ratio
between the functional determinant coming from the ghosts\footnote{
The gauge fixing \rf{singular} requires by \rf{topo}
a Faddeev--Popov term $\bar q \wedge
d_Aq$, where $\bar q$ is the anti-self-dual
anti-ghost associated to the 1-form ghost $q$.
}
and
the Jacobian associated with the delta-functional.

\section{Donaldson's polynomial invariants}
\label{sezsei}

In this section we show how the original Donaldson polynomial
invariants~\cite{dk,w2} naturally come in the
framework of our theory after the introduction of Wilson-line
operators
\beq\label{wilson}
W(C)=\tr P\exp ( \oint_C A)
\eeq
Let us now consider a closed oriented surface
$\hat\Sigma$. We open a small contractible disk $\Sigma''$ so that
$\hat\Sigma=\Sigma'\cup\Sigma''$
and compute the following expectation value:
\beq\label{ev}
\la M(\Sigma)W(C)\ra
\eeq
where $C=\partial\Sigma'=-\partial \Sigma''$.
Since the Wilson loop does not depend on the $B$-field,
we can perform the $B$ integration exactly as in the previous
section.
Then the resulting delta-functional restricts the $A$-integration
to the reducible connections $\tilde A$  and this allows to write
the Wilson loop as
\beq\label{wilsonuno}
W(C)=\tr\exp (\int_{\Sigma'} f \, t_3)
\eeq
Since in this case one has
$f=\mp 2\pi q_mg\omega[\Sigma]$
and
$2gq_m Q(\Sigma,\Sigma)=1+4n$,~\rf{ev} gives
\beq\label{QQ}
\la M(\Sigma)W(C)\ra=\la M(\Sigma)\ra
2\cos\left[
{\pi\over 2}(1+4n){Q(\Sigma,\hat\Sigma)\over Q(\Sigma,\Sigma)}
\right]
\eeq
If the monopole has the fundamental charge ($n=0$),~\rf{QQ}
gives the algebraic intersection number $Q(\Sigma,\hat\Sigma)$
modulo $2Q(\Sigma,\Sigma)$. Notice that the insertion of the operator
$M(\Sigma)$ creates a monopole whose flux through $\Sigma'$ measures
the intersection number.

We now move to the general case of $d+\hat d$ classes
$[\Sigma_1],\ldots,[\Sigma_d]$
and $[\hat\Sigma_1],\ldots,[\hat\Sigma_{\hat d}]$
in $H_2(X,{\bf Z})$.
We can associate as before $\hat d$ loops $C_1,\ldots,C_{\hat d}$
to the $[\hat\Sigma_b]$ and compute the expectation value
\beq\label{eevv}
\la \prod_{a=1}^d M(\Sigma_a)\prod_{b=1}^{\hat d}
W(C_b)\ra
\eeq
which turns out to be proportional to
\beq\label{vvee}
2^{\hat d}\prod_{a=1}^d\prod_{b=1}^{\hat d}
\cos\left(
{\pi\over 2}(1+4n_a){Q(\Sigma_a,\hat\Sigma_b)\over Q_0(\Sigma_.)}
\right)
\eeq
where
$Q_0(\Sigma_.)={\rm G.C.D. } \{Q(\Sigma_a,\Sigma_a); a=1,\ldots,d\}$.

Moreover, choosing $d=\hat d$ and $\Sigma_a=\hat\Sigma_a$,
$a=1,\ldots ,d$, one gets a symmetrized version of~\rf{vvee},
which is a generating functional for the Donaldson polynomial invariants
associated to $X$,
$q_X([\Sigma_1],\ldots,[\Sigma_d])$
\cite{w2}.

{}From a physical point of view the Wilson-loops
detect the monopole flux generated by
the insertion of the operators  $M(\Sigma)$ associated to surfaces.
Monopoles of the same charge on the two sides of the
surfaces give opposite contributions to the
flux\rlap,\footnote{
This picture, usually understood as a
``double monopole layer'', has been suggested to us by A.~Sagnotti
and has been firstly considered by A.~Polyakov~\cite{pol}
in 3D-QED and more recently in Ref.~\cite{msz} in
4D-QED.
}
which turn out to be proportional to the algebraic intersection numbers.

\section{Conclusions}
In this work we have studied
the self-dual $BF$ theory as the first order version
of pure Yang--Mills theory with theta-term.
In this framework
we have the Wilson loop operator $W(C)$, which
is the ``order operator" for ordinary Yang--Mills theory,
and an operator $M(\Sigma)$ playing the r\^ole of
a ``disorder parameter'' in the sense
of Ref.~\cite{th1}. Indeed
the abelian reduction \rf{ventinovea} maps the highly non-trivial relations
among the ``Weyl group operators"\footnote{
Notice that by \rf{BF} in $3+1$ dimensions the spatial components
$\epsilon_{ijk}B_{jk}$ are canonically conjugated to $A_i$.
}
$\rm Hol$ and $M$ into the 't~Hooft commutation relations.

{}From a physical point of view, the insertion of $M(\Sigma)$
plays the r\^ole of the standard ``abelian projection'' in pure QCD
without Higgs mechanism~\cite{th2}.

The  expectation  values  of
$M(\Sigma)$ alone and of the combination $M(\Sigma)W(C)$ are related
to the Donaldson--Witten invariants of
a K\"ahler manifold $X$
\cite{dk,w3,w2,w1,t,km}.
However, these invariants arise here in a different setting with respect to
the one considered by Witten.

Yet the computation of the invariants relies on the
existence of monopole equations
both here and in Witten's approach.

One of the differences between our framework and Witten's one,
is that our observables introduce singularities, so our theory
appears to be non-trivial, even if we work with ${\bf R}^4$.
For a related approach see also \cite{ans}.

The observable $M(\Sigma)$ (in the case when $\Sigma$ is a torus)
can be given a geometrical interpretation as parallel transport
of one of the fundamental cycles along the other. This suggests that $BF$
theory can be described as a ``gauge theory of loops"
\cite{ccmr}.
Our formulation gives an effective field-theoretical
description of the string picture for QCD introduced in Ref.~\cite{p}.

All our calculations have been done in
the limit $g\to 0$. In this limit both the topological
$BF$ theory with the
``cosmological term" \rf{BF} in 4-dimensions  and the Yang--Mills theory with
theta-term appear to detect 4-manifolds invariants. For these two theories
to be equivalent (on shell),  the choice of the ``gauge" $B^- =0$
(that breaks the larger ``topological invariance" of \rf{BF}) is
a key ingredient.

If we instead consider a ``pure" $BF$ theory (i.e.\ we take $g=0$) and do
not impose the constraint of self-duality $B^-=0$,
we can study a different topological field theory,
as discussed in \cite{cm,c,ccfm}.
Perturbation theory in  $\kappa=q_mg$
detects 2-knots, i.e. embedded (or immersed) 2-surfaces, up to
diffeomorphisms of the 4-manifolds.

\begin{center}
{\bf Acknowledgements}
\end{center}
\noindent
The authors acknowledge  fruitful  discussions  with J.~Baez,
M.~Bianchi, J.~Fr\"ohlich,  F.~Fucito,  G.~Rossi,
A.~Sagnotti and M.~Zeni.
In particular we thank R.~Catenacci for some critical remarks.

\end{document}